\documentclass[aps,prd,twocolumn,nofootinbib]{revtex4}

\usepackage{graphicx}
\usepackage{hyperref}
\usepackage{slashed}
\usepackage{color}

\begin{document}

\title {Excited heavy quarkonium production via $Z^0$ decays at a high luminosity collider}

\author{Qi-Li Liao$^{1}$}
\email{xiaosueer@163.com}
\author{Yan Yu$^{1}$}
\author{Ya Deng$^{1}$}
\author{Guo-Ya Xie$^{2}$}
\author{Guang-Chuan Wang$^{1}$}
\address{$^{1}$College of Mobile Telecommunications Chongqing  University of Posts and Telecom, Chongqing 401520, P.R. China\\$^{2}$ Chongqing  University of Posts and Telecom, Chongqing 400065, P.R. China}

\date{\today}
\begin{abstract}
We present a systematic study of the production of the heavy quarkonium, i.e., $|(c\bar{c})[n] \rangle$ , $|(b\bar{c})[n] \rangle$  (or $|(c\bar{b})[n] \rangle$),  and $|(b\bar{b})[n] \rangle$ quarkonium [$|(Q\bar{Q'})[n]\rangle$ quarkonium for short], through $Z^0$ boson semiexclusive decays with new parameters \cite{lx} for the heavy quarkonium under the framework of the nonrelativistic QCD, where $[n]$ stands for $n^1S_0$, $n^3S_1$, $n^1P_0$, $n^3P_J$ ($n=1, \cdots, 6$; $J=[0, 1, 2]$). ``Improved trace technology" is adopted to derive the simplified analytic expressions at the amplitude level, which shall be useful for dealing with these decay channels. If all higher $|(Q\bar{Q'})[n]\rangle$ quarkonium states decay to the ground state $|(Q\bar{Q'})[1^1S_0]\rangle$ with $100\%$ efficiency via electromagnetic or hadronic interactions, we obtain $\Gamma{(Z^0\to |(c\bar{c})[1^1S_0]\rangle)}=1476$ KeV, $\Gamma{(Z^0\to |(b\bar{c})[1^1S_0]\rangle)}=1485$ KeV, $\Gamma{(Z^0\to |(b\bar{b})[1^1S_0]\rangle)}=127.5$ KeV. At the LHC and ILC with the luminosity ${\cal L}\propto 10^{34}cm^{-2}s^{-1}$, sizable heavy quarkonium events can be produced through $Z^0$ boson decays; i.e., about $5.9~\times10^{5}$ $(c\bar{c})$, $6.0~\times10^{5}$ $(b\bar{c})$ [or $(c\bar{b})$], $5.1~\times10^{4}$ $(b\bar{b})$ events per year can be obtained.\\


\end{abstract}

\maketitle

\section{Introduction}

The impressive volume of experimental data on multiple heavy quarkonium production were obtained by the experiments at the LHC in the last years. The LHCb and CMS Collaboration experiments have published studies of the $B_c$-meson production and of the double $J/\Psi$ production \cite{r,cc}. Also very interesting is that the candidate of the $B_c$(2S) state has been observed by ATLAS \cite{ga}.

At the LHC \cite{sc}, the International Linear Collider (ILC) \cite{jsw,g}, and the newly purposed $Z$ factory \cite{mz}, the $Z^0$ boson is being and will be copiously produced. The $Z^0$ boson production cross section is about $34$nb at the LHC with collider runs at the center-of-mass energy $\sqrt{S}=14$ TeV and will be about $30$nb at the ILC runs at the $Z^0$ pole energy \cite{jsw}. There will be about $10^9$ $Z^0$ boson events being produced per year at the LHC and ILC with the luminosity ${\cal L}\propto 10^{34}cm^{-2}s^{-1}$. Therefore, investigation of the heavy quarkonium production through $Z^0$ decays is worthwhile and meaningful. The study of the heavy quarkonium, e.g., $|(c\bar{c})[n] \rangle$, $|(b\bar{c})[n] \rangle$ (or $|(c\bar{b})[n] \rangle$),  and $|(b\bar{b})[n] \rangle$ quarkonium, can help us to achieve a deeper understanding of the quantum chromodynamics (QCD) in both the perturbative and nonperturbative sectors. The heavy quarkonium is presumed to be a nonrelativistic bound state of the heavy quark and antiquark. A wonderful theoretical tool to deal with the processes involving quarkonium is the nonrelativistic quantum chromodynamics (NRQCD) \cite{nrqcd}, in which the low-energy interactions are organized by the expansion in $v$, where $v$ stands for the typical relative velocity of the heavy quark and antiquark inside of the heavy quarkonium. The heavy quarkonium production itself is very useful for high precision physics in the electroweak sector and testing the perturbative QCD (pQCD) \cite{nb1,nb2,glb}. For example, since its discovery by the CDF Collaboration \cite{fa}, the $B_c$ meson being the unique ``doubly heavy-flavored" meson in the standard model (SM) has aroused great interest. The direct hadronic production of the $B_c$ meson has been studied systematically in Refs. \cite{kar,spb,cyg,aam,cjx,ccpx}. For compensation, it would be helpful to study its indirect production mechanisms. A systematic study of the $B_c$ meson production through the $t$ quark or $\bar{t}$ quark, $W\pm$, and the $Z^0$  boson decay can be found in the literature \cite{lx,cck,w,tbc2,zbc0,zbc1,zbc2,wbc1,wbc2}. Because of sizable $t/\bar{t}$ quark, $W^\pm$ and $Z^0$ boson events shall be produced at the LHC and ILC, the production of $B_c$ through their decay shall be helpful for determining the $B_c$ meson properties.

Due to a high collision energy and high luminosity at the LHC and ILC, sizable amounts of the heavy quarkonium events can be produced through $Z^0$ decays \cite{jsw,g}. So these channels may be an important supplement for other measurements at the LHC and ILC. Here we will further discuss the production of even higher $|(Q\bar{Q'})[n]\rangle$ quarkonium Fock states through the $Z^0$ semiexclusive decays with new parameters \cite{lx}, where $|(Q\bar{Q'})[n]\rangle$ stands for $|(Q\bar{Q'})[n^1S_0]\rangle$, $|(Q\bar{Q'})[n^3S_1]\rangle$, $|(Q\bar{Q'})[n^1P_1]\rangle$, $|(Q\bar{Q'})[n^3P_J]\rangle$ quarkonium [with $n=1, \cdots, 6$; $J=(0 , 1, 2)$], respectively, and $Q$ or $Q'$ is $c(\bar{c}$) or $b(\bar{b}$) quark, respectively. As shown in Refs. \cite{wbc1,sun,zbc2}, due to the color suppression of the amplitude and the relative-velocity suppression of the color-octet matrix element, the color-octet $|(Q\bar{Q'})g[n]\rangle$ component provides negligible contributions. We will only discuss the color-singlet state¡¯ production channels.

Under the NRQCD framework \cite{nrqcd}, a doubly heavy quarkonium is considered as an expansion of various Fock states. The relative importance among those infinite ingredients is evaluated by the velocity scaling rule. To deal with heavy $|(Q\bar{Q'})[n]\rangle$ quarkonium production through $Z^0$ semiexclusive decays, one needs to derive the pQCD calculable squared amplitudes. But the analytical expression for the usual squared amplitude $|\Sigma|^2$ becomes too complex and lengthy for more (massive) particles in the final states and for higher-level Fock states to be generated for the emergence of massive-fermion lines in the Feynman diagrams, especially to derive the amplitudes of the $P$-wave states. On account of needing to get the derivative of the amplitudes over the relative momentum of the constituent quarks for the $P$-wave states. It has been found that to do the numerical calculation using this conventional squared-amplitude technology becomes time-consuming for these complex processes, since the cross terms of the matrix elements increase with the increment of Feynman diagrams $|\Sigma|^2=\Sigma_{ij}M_iM^*_j$, where $i$ and $j$ stand for the number of Feynman diagrams of the process. To solve this problem, the ``improved trace technology" is suggested and developed in the literature ~\cite{lx,tbc2,zbc0,zbc1,zbc2,wbc1,wbc2}; it deals with the process directly at the amplitude level. After generating proper phase-space points, one first calculates the numerical value for the amplitudes, sums these values algebraically, and then squares the sum to get the squared amplitude, $|\Sigma|^2=|\Sigma_{i}M_i|^2$; through this method, numerical simulation efficiency can be greatly improved in comparison to the conventional squared-amplitude technology. Moreover, under this approach, many simplifications can be done at the amplitude level due to the fermion-line symmetries and the specific properties of each heavy quarkonium Fock state; then, one can even write down the analytic expressions for the amplitude. We will continue to adopt improved trace technology to derive the analytical expression for all the above-mentioned decay channels.

The rest of the paper is organized as follows. In Sec. II, we introduce the calculation techniques for the $Z^0$ boson semiexclusive decays to $|(Q\bar{Q'})[n]\rangle$ quarkonium, where $[n]$ stands for $n^1S_0$, $n^3S_1$, $n^1P_0$, $n^3P_J$ ($n=1,\cdots,6$; $J=[0, 1, 2]$). Then we calculate the production of $|(c\bar{c})[n]\rangle$, $|(b\bar{c})[n]\rangle$, and $|(b\bar{b})[n]\rangle$ quarkonium through $Z^0$ decay channels, i.e. $Z^0\to |(c\bar{c})[n]\rangle+\bar{c}c$, $Z^0\to |(b\bar{c})[n]\rangle+\bar{c}b$, and $Z^0\to |(b\bar{b})[n]\rangle+\bar{b}b$, with new parameters \cite{lx} for the $|(Q\bar{Q'})[n]\rangle$ quarkonium in Sec. III. In Sec. IV, we discuss the production of $|(Q\bar{Q'})[n]\rangle$ quarkonium via $Z^0$ boson decays with the different number of flavor quarks $n_f$ under the five potential models. The final section is reserved for a summary.

\section{CALCULATION TECHNIQUES AND FORMULATION}

\begin{figure}
\includegraphics[width=0.45\textwidth]{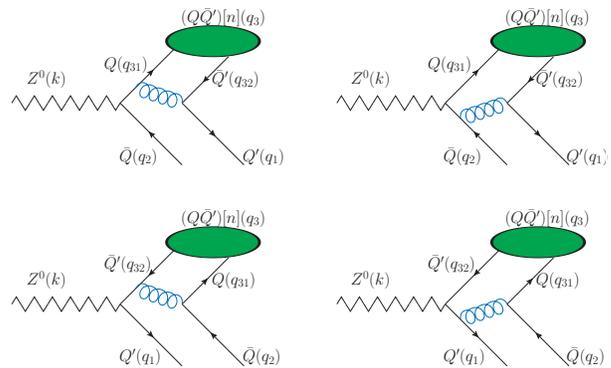}
\caption{(color online). Feynman diagrams for the process $Z^0(k)\rightarrow |(Q\bar{Q'})[n]\rangle(q_3) + \bar{Q}(q_2)+Q'(q_1)$, where $|(Q\bar{Q'})[n]\rangle$ stands for $|(Q\bar{Q'})[n^1S_0]\rangle$, $|(Q\bar{Q'})[n^3S_1]\rangle$, $|(Q\bar{Q'})[n^1P_1]\rangle$, $|(Q\bar{Q'})[n^3P_J]\rangle$ quarkonium (with $n=1, \cdots, 6$; $J=(0 , 1, 2)$), respectively. $Q$ or $Q'$ is $c(\bar{c}$) or $b(\bar{b}$) quark, respectively.} \label{feyn1}
\end{figure}

The four semiexclusive processes for the heavy quarkonium production through $Z^0$ boson decays can be dealt with, i.e., $Z^0\to |(c\bar{c})[n]\rangle+\bar{c}c$, $Z^0\to |(b\bar{c})[n]\rangle+\bar{b}c$ (or $Z^0\to |(c\bar{b})[n]\rangle+\bar{c}b$), and $Z^0\to |(b\bar{b})[n]\rangle+\bar{b}b$. These processes can be shorten for $Z^0 (k) \to |(Q\bar{Q'})[n]\rangle(q_3) +\bar{Q}(q_2) + Q'(q_1)$, where $Q$ or $Q'$ stands for charm or beauty quark, and $k$ and $q_i$ ($i=1, 2, 3$) are the momenta of the corresponding particles. According to the NRQCD factorization formula \cite{nrqcd}, the square of the semiexclusive amplitude can be written as the production of the perturbatively calculable short-distance coefficients and the nonperturbative long-distance factors, the so-called nonperturbative NRQCD matrix elements. Its total decay widths $d\Gamma$ can be factorized as

\begin{equation}
d\Gamma=\sum_{n} d\hat\Gamma(Z^0\to |(Q\bar{Q'})[n]\rangle+ \bar{Q}Q') \langle{\cal O}^H(n) \rangle,
\end{equation}
where $\langle{\cal O}^{H}(n)\rangle$ is the nonperturbative matrix element which describes the hadronization of a $Q\bar{Q'}$ pair into the observable quark state $H$ and is proportional to the transition probability of the perturbative state $Q\bar{Q'}$ into the bound state $|(Q\bar{Q'})[n]\rangle$. As for the color-singlet components, the nonperturbative matrix elements can be directly related either to the wave functions at the origin or the first derivative of the wave functions at the origin \cite{nrqcd}, which can be computed via the potential models \cite{lx,pot1,pot2,pot3,pot4,pot5} and/or potential NRQCD (pNRQCD) \cite{pnrqcd1,nb2,yellow} and/or lattice QCD \cite{lat1}.

The short-distance decay width $\hat\Gamma(Z^0)$ can be expressed as

\begin{equation}
d\hat\Gamma(Z^0\to |(Q\bar{Q'})[n]\rangle+\bar{Q}Q')= \frac{1}{2 k^0} \overline{\sum}  |M|^{2} d\Phi_3,
\end{equation}
where $\overline{\sum}$ means that one needs to average over the spin states of the initial particle and to sum over the color and spin of all the final particles. In the $Z^0$ rest frame, the three-particle phase space can be written as
\begin{equation}
d{\Phi_3}=(2\pi)^4 \delta^{4}\left(k_{0} - \sum_f^3 q_{f}\right)\prod_{f=1}^3 \frac{d^3{\vec{q}_f}}{(2\pi)^3 2q_f^0}.
\end{equation}
The process to simplify the $1 \to 3$ phase space with a massive quark/antiqark in the final state has been dealt with in greater detail in Refs.~\cite{tbc2,zbc1}. To shorten the paper, we shall not present it here, but the interested reader may turn to these references for the detailed technology. With the help of the formulas listed in Refs.~\cite{tbc2,zbc1}, one can not only derive the whole decay widths but also obtain the corresponding differential decay widths that are helpful for experimental studies, such as $d\Gamma/ds_{1}$, $d\Gamma/ds_{2}$, $d\Gamma/d\cos\theta_{12}$, and $d\Gamma/d\cos\theta_{13}$, where $s_{1}=(q_1+q_2)^2$, $s_{2}=(q_1+q_3)^2$, $\theta_{12}$ is the angle between $\vec{q}_1$ and $\vec{q}_2$, and $\theta_{13}$ is the angle between $\vec{q}_1$ and $\vec{q}_3$.

The hard-scattering amplitude for the specified processes can be dealt with as follows:
\begin{eqnarray}
&&Z^0\rightarrow |(c\bar{c})[n]\rangle + \bar{c}c,~~~Z^0\rightarrow |(b\bar{c})[n]\rangle + \bar{b}c,\nonumber\\
&&Z^0\rightarrow |(b\bar{b})[n]\rangle + \bar{b}b.
\end{eqnarray}

The Feynman diagrams of the above three processes as $Z^0 (k) \to |(Q\bar{Q'})[n]\rangle(q_3) +\bar{Q}(q_2) + Q'(q_1)$ is presented in Fig.~\ref{feyn1}, where the intermediate gluon should be hard enough to produce a $b\bar{b}$ pair or $c\bar{c}$ pair, so the amplitude is pQCD calculable.

These amplitudes can be generally expressed as
\begin{equation} \label{amplitude}
i{\cal{M} }= {\cal{C}} {\bar {u}_{s i}}({q_2}) \sum\limits_{n = 1}^{m} {{\cal A} _n } {v_{s' j}}({q_1}),
\end{equation}
where $m$ stands for the number of Feynman diagrams, $s$ and $s'$ are spin states, and $i$ and $j$ are color indices for the outing $Q$, $Q'$ quark. The overall factor $\cal{C}$ stands for the specified quarkonium in the color-singlet, where ${\cal C}=\frac{g g^2_s}{4~cos\theta_W}\times \frac{4}{3\sqrt{3}} \delta_{ij}$, $\theta_W$ stands for the Weinberg angle. ${\cal A} _n $ in the formulas is similar in Ref.~\cite{zbc2}, so do not listed here to shorten the paper.

By using the improved trace technology, one can sequentially obtain the squared amplitudes, and the numerical efficiency can also be greatly improved ~\cite{lx,tbc2,zbc0,zbc1,zbc2,wbc1,wbc2}. We adopt the improved trace technology to simplify the amplitudes $M_{ss^{\prime}}$ at the amplitude level for the above-mentioned processes, the standard procedures for $Z^0(k) \rightarrow |(Q\bar{Q'})[n]\rangle(q_3) + \bar{Q}(q_2)+Q'(q_1)$ are similarly presented in Ref.~\cite{zbc2}.

Finally, the decay widths over $s_{1}$ and $s_{2}$ can be expressed as
\begin{equation}
d\Gamma= \frac{\langle{\cal O}^H(n) \rangle}{256 \pi^3 m^3_Z}( \overline{\sum}|M|^{2}) ds_{1}ds_{2},
\end{equation}
where $m_Z$ is the mass of the $Z^0$ boson. The color-singlet nonperturbative matrix element $\langle{\cal O}^H(n) \rangle$ can be related either to the Schr${\rm \ddot{o}}$dinger wave function $\psi_{(Q\bar{Q'})}(0)$ at the origin for the $S$-wave quarkonium states or the first derivative of the wave function $\psi^\prime_{(Q\bar{Q'})}(0)$ at the origin for the $P$-wave quarkonium states:
\begin{eqnarray}
\langle{\cal O}^H(nS) \rangle &\simeq& |\psi_{\mid(Q\bar{Q'})[nS]\rangle}(0)|^2,\nonumber\\
\langle{\cal O}^H(nP) \rangle &\simeq& |\psi^\prime_{\mid(Q\bar{Q'})[nP]\rangle}(0)|^2.
\end{eqnarray}

Since the spin-splitting effects are small, the difference between the wave function parameters for the spin-singlet and spin-triplet states at the same level are not distinguished. The Schr${\rm \ddot{o}}$dinger wave function at the origin $\Psi_{|Q\bar{Q'})[nS]\rangle}(0)$ and the first derivative of the Schr${\rm \ddot{o}}$dinger wave function at the origin $\Psi^{'}_{|(Q\bar{Q'})[nP]\rangle}(0)$ are related to the radial wave function at the origin $R_{|(Q\bar{Q'})[nS]\rangle}(0)$ and the first derivative of the radial wave function at the origin $R^{'}_{|(Q\bar{Q'})[nP]\rangle}(0)$, respectively,
\begin{eqnarray}
\Psi_{|(Q\bar{Q'})[nS]\rangle}(0)&=&\sqrt{{1}/{4\pi}}R_{|(Q\bar{Q'})[nS]\rangle}(0),\nonumber\\
\Psi'_{|(Q\bar{Q'})[nP]\rangle}(0)&=&\sqrt{{3}/{4\pi}}R'_{|(Q\bar{Q'})[nP]\rangle}(0).
\end{eqnarray}

The radial wave function at the origin $R_{|(Q\bar{Q'})[nS]\rangle}(0)$ and the first derivative of the radial wave function at the origin $R^{'}_{|(Q\bar{Q'})[nP]\rangle}(0)$ relate to the number of active flavor quarks $n_f$, the constituent quark mass of the $|(Q\bar{Q}')[n]\rangle$ quarkonium, and the concrete potential models, respectively~\cite{lx}. Thus, $R_{|(Q\bar{Q'})[nS]\rangle}(0)$ and $R^{'}_{|(Q\bar{Q'})[nP]\rangle}(0)$ are adopted in Refs.~\cite{lx}.

\section{Numerical Results}

\subsection{Input parameters}

The input parameters are adopted as the following values \cite{wtd,pdg}: $m_Z =91.1876$ GeV, $Z^0$ full width $\Gamma(Z^0)=2.4952$ GeV. $\theta_W=\arcsin\sqrt{0.23119}$ is the Weinberg angle, $m_{W}=80.399$GeV. We set the renormalization scale to be $m_{(c\bar{c})}$ and $m_{(b\bar{c})}$ of $|(c\bar{c})\rangle$ and $|(b\bar{c})\rangle$ quarkonium for leading-order $\alpha_s$ running , which leads to $\alpha_s=0.26$ and $m_{(b\bar{b})}$ of $|(b\bar{b})\rangle$-quarkonium for $\alpha_s=0.18$. To ensure the gauge invariance of the hard amplitude, we set the $|(Q\bar{Q'})[n]\rangle$ quarkonium mass $M$ to be $m_Q+m_{Q'}$. We adopt the values derived in Refs.~\cite{lx,pdg} and list them in Table~\ref{tabrpa}, since it is noted that the Buchm${\rm \ddot{u}}$ller and Tye potential (B.T. potential) has the correct two-loop short-distance behavior in QCD~\cite{pot2,wgs} and the decay widths are related to the constituent quark mass of the $|(Q\bar{Q}')[n]\rangle$ quarkonium.

\subsection{Heavy quarkonium production via $Z^0$ decays}

The decay widths for the $|(Q\bar{Q'})[n]\rangle$ quarkonium states and the production channels through $Z^0$ decays, i.e., $Z^0\rightarrow |(c\bar{c})[n]\rangle+\bar{c}c$, $Z^0\rightarrow |(b\bar{c})[n]\rangle+\bar{b}c$ ,and $Z^0\rightarrow |(b\bar{b})[n]\rangle+\bar{b}b$, are listed in Tables \ref{tabrpb}, \ref{tabrpc}, and \ref{tabrpd} within the B.T. potential \cite{lx}.

\begin{widetext}
\begin{center}
\begin{table}
\caption{Mass of the constituent quark and radial wave functions at the origin under the B.T. potential \cite{lx}.}
\begin{tabular}{|c||c|c|c|c|c|c|c|c|}
\hline\hline
$|(Q\bar{Q'})[n]\rangle$&Mass and wave functions&~$n=1$~~&~$n=2$~~&~$n=3$~~&~$n=4$~~&~$n=5$~~&~$n=6$~~&~$n=7$~~\\
\hline
$S$ states of $|(c\bar{c})[n]\rangle$&$m_{c}~({GeV})$&1.48&1.82&1.92&2.02&2.12&2.25&~~~\\
~~~&$|R_{|[nS]\rangle}(0)|^2({GeV}^3)$~($n_{f}$=3)&~2.458&~1.617&~0.969&~0.796&~0.701&~0.721&~~~\\
\hline
$P$ states of $|(c\bar{c})[n]\rangle$&$m_{c}~({GeV})$&1.75&1.96&2.12&2.26&2.38&~~~&~~~\\
~~~~&$|R'_{|[np]\rangle}(0)|^2({GeV}^5)$~($n_{f}$=3)&~0.322&~0.224&~0.387&~0.467&~0.499&~~~&~~~\\
\hline
$S$ states of $|(b\bar{c})[n]\rangle$&$m_{c}~({GeV})$&1.45&1.82&1.96&2.10&2.15&~~~&~~~\\
~~~~&$m_{b}~({GeV})$&4.85&5.03&5.15&5.30&5.45&~~~&~~~\\
~~~&$|R_{|[nS]\rangle}(0)|^2({GeV}^3)$~($n_{f}$=3)&~3.848&~1.987&~1.347&~1.279&~1.118&~~~&~~~\\
\hline
$P$ states of $|(b\bar{c})[n]\rangle$&$m_{c}~({GeV})$&1.75&1.96&2.15&2.26&~~~&~~~&~~~\\
~~~&$m_{b}~({GeV})$&4.93&5.13&5.25&5.37&~~~&~~~&~~~\\
~~~&$|R'_{|[np]\rangle}(0)|^2({GeV}^5)$~($n_{f}$=3)&~0.518&~0.500&~0.729&~0.823&~~~~&~~~&~~~\\
\hline
$S$ states of $|(b\bar{b})[n]\rangle$&$m_{b}~({GeV})$&4.71&5.01&5.17&5.27&5.41&5.50&5.58\\
~~~&$|R_{|[nS]\rangle}(0)|^2({GeV}^3)$~($n_{f}$=4)&~16.12&~6.746&~2.172&~2.588&~2.665&~2.576&~2.377\\
\hline
$P$ states of $|(b\bar{b})[n]\rangle$&$m_{b}~({GeV})$&4.94&5.12&5.20&5.37&5.47&5.56&~~ \\
~~~&$|R'_{|[np]\rangle}(0)|^2({GeV}^5)$~($n_{f}$=4)&~5.874 &~2.827&~2.578&~3.217&~3.573&~3.669&~~\\
\hline\hline
\end{tabular}
\label{tabrpa}
\end{table}

\begin{table}
\caption{Decay widths (in unit KeV) for the production of $|(c\bar{c})[n]\rangle$ ($n_{f}=3$) through $Z^0$ decays.}
\begin{tabular}{|c||c|c|c|c|c|c|c|c|}
\hline\hline
~~~~ & ~~$n=1$~~ & ~~$n=2$&~~ $n=3$&~~ $n=4$&~~ $n=5$&~~ $n=6$\\
\hline
$Z^0\rightarrow (c\bar{c})[n^1S_0] +\bar{c}c$ &~~313.0~~&~~108.0~~&~~54.67~~ &~~38.25~~&~~28.90~~&~~24.59~~\\
\hline
$Z^0\rightarrow (c\bar{c})[n^3S_1] +\bar{c}c$&~~325.0~~&~~112.3~~&~~56.93~~ &~~39.86~~&~~30.13~~&~~25.66~~\\
\hline
$Z^0\rightarrow (c\bar{c})[n^1P_1] +\bar{c}c$&~~27.84~~&~~11.66~~&~~11.91~~ &~~10.50~~&~~8.473~~&~~~~\\
\hline
$Z^0\rightarrow (c\bar{c})[n^3P_0] +\bar{c}c$&~~39.15~~&~~14.81~~&~~17.43~~ &~~15.12~~&~~12.49~~&~~~~\\
\hline
$Z^0\rightarrow (c\bar{c})[n^3P_1] +\bar{c}c$&~~41.52~~&~~16.22~~&~~18.78~~ &~~16.34~~&~~13.40~~&~~~~\\
\hline
$Z^0\rightarrow (c\bar{c})[n^3P_2] +\bar{c}c$&~~18.32~~&~~5.857~~&~~7.173~~ &~~6.303~~&~~5.380~~&~~~~\\
\hline\hline
\end{tabular}
\label{tabrpb}
\end{table}

\begin{table}
\caption{Decay widths (in unit KeV) for the production of $|(b\bar{c})[n]\rangle$ ($n_{f}=3$) through $Z^0$ decays.}
\begin{tabular}{|c||c|c|c|c|c|c|c|c|}
\hline\hline
~~~~ & ~~$n=1$~~ & ~~$n=2$&~~ $n=3$&~~ $n=4$&~~ $n=5$\\
\hline
$Z^0\rightarrow (b\bar{c})[n^1S_0] +\bar{b}c$ &~~342.2~~&~~87.71~~&~~47.22~~ &~~36.11~~&~~29.22~~\\
\hline
$Z^0\rightarrow (b\bar{c})[n^3S_1] +\bar{b}c$&~~492.4~~&~~119.5~~&~~63.46~~ &~~48.02~~&~~38.96~~\\
\hline
$Z^0\rightarrow (b\bar{c})[n^1P_1] +\bar{b}c$&~~17.09~~&~~9.237~~&~~8.617~~ &~~7.592~~&~~~~\\
\hline
$Z^0\rightarrow (b\bar{c})[n^3P_0] +\bar{b}c$&~~11.69~~&~~6.962~~&~~6.954~~ &~~6.313~~&~~~~\\
\hline
$Z^0\rightarrow (b\bar{c})[n^3P_1] +\bar{b}c$&~~21.41~~&~~12.05~~&~~11.41~~ &~~10.14~~&~~~~\\
\hline
$Z^0\rightarrow (b\bar{c})[n^3P_2] +\bar{b}c$&~~20.97~~&~~11.20~~&~~9.954~~ &~~8.651~~&~~~~\\
\hline\hline
\end{tabular}
\label{tabrpc}
\end{table}

\begin{table}
\caption{Decay widths (in unit KeV) for the production of $|(b\bar{b})[n]\rangle$ ($n_{f}=4$) through $Z^0$ decays.}
\begin{tabular}{|c||c|c|c|c|c|c|c|c|}
\hline\hline
~~~~ & ~~$n=1$~~ & ~~$n=2$&~~ $n=3$&~~ $n=4$&~~ $n=5$&~~ $n=6$&~~ $n=7$\\
\hline
$Z^0\rightarrow (b\bar{b})[n^1S_0] +\bar{b}b$ &~~28.48~~&~~9.537~~&~~2.737~~ &~~3.039~~&~~2.839~~&~~2.581~~&~~2.256~~\\
\hline
$Z^0\rightarrow (b\bar{b})[n^3S_1] +\bar{b}b$&~~31.83~~&~~10.74~~&~~3.095~~ &~~3.446~~&~~3.231~~&~~2.944~~&~~2.579~~\\
\hline
$Z^0\rightarrow (b\bar{b})[n^1P_1] +\bar{b}b$&~~1.387~~&~~0.550~~&~~0.461~~ &~~0.483~~&~~0.484~~&~~0.455~~&~~~~\\
\hline
$Z^0\rightarrow (b\bar{b})[n^3P_0] +\bar{b}b$&~~2.199~~&~~0.878~~&~~0.739~~ &~~0.779~~&~~0.785~~&~~0.740~~&~~~~\\
\hline
$Z^0\rightarrow (b\bar{b})[n^3P_1] +\bar{b}b$&~~2.104~~&~~0.835~~&~~0.700~~ &~~0.734~~&~~0.738~~&~~0.693~~&~~~~\\
\hline
$Z^0\rightarrow (b\bar{b})[n^3P_2] +\bar{b}b$&~~0.870~~&~~0.345~~&~~0.290~~ &~~0.304~~&~~0.305~~&~~0.287~~&~~~~\\
\hline\hline
\end{tabular}
\label{tabrpd}
\end{table}
\end{center}
\end{widetext}

From Tables~\ref{tabrpb}-\ref{tabrpd}, it is noted that, in addition to the ground $1S$-level states, the higher $|(Q\bar{Q'})[n]\rangle$ quarkonium states can also provide sizable contributions to the total decay widths. For convenience, we have used $[nS]$ to present the summed decay widths of $[n^1S_0]$ and $[n^3S_1]$ at the same $n$th level, and $[nP]$ to represent the summed decay widths of $[n^1P_1]$ and $[n^3P_J] (J=0, 1, 2)$ at the same $n$th level.

\begin{widetext}
\begin{center}
\begin{figure}
\includegraphics[width=0.40\textwidth]{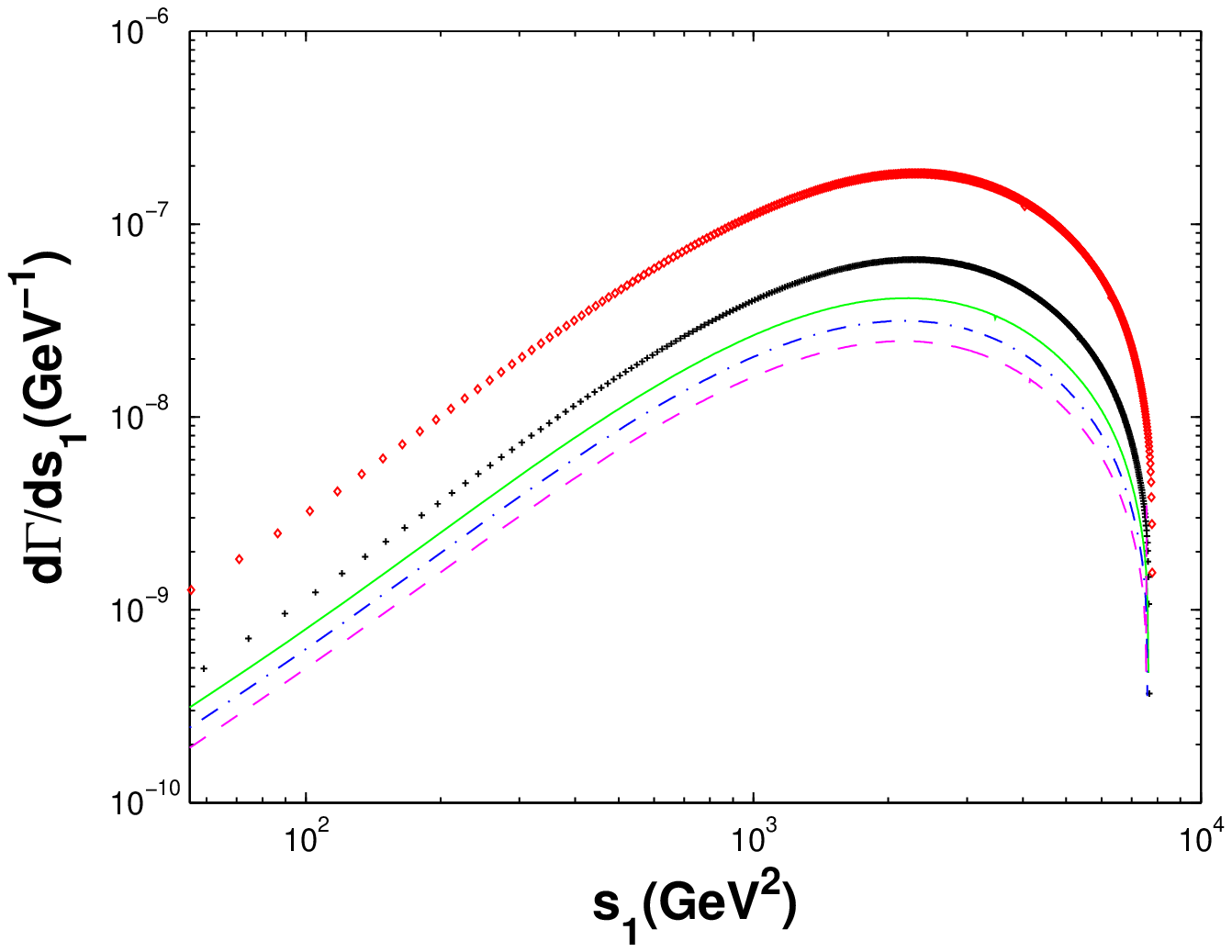}
\includegraphics[width=0.40\textwidth]{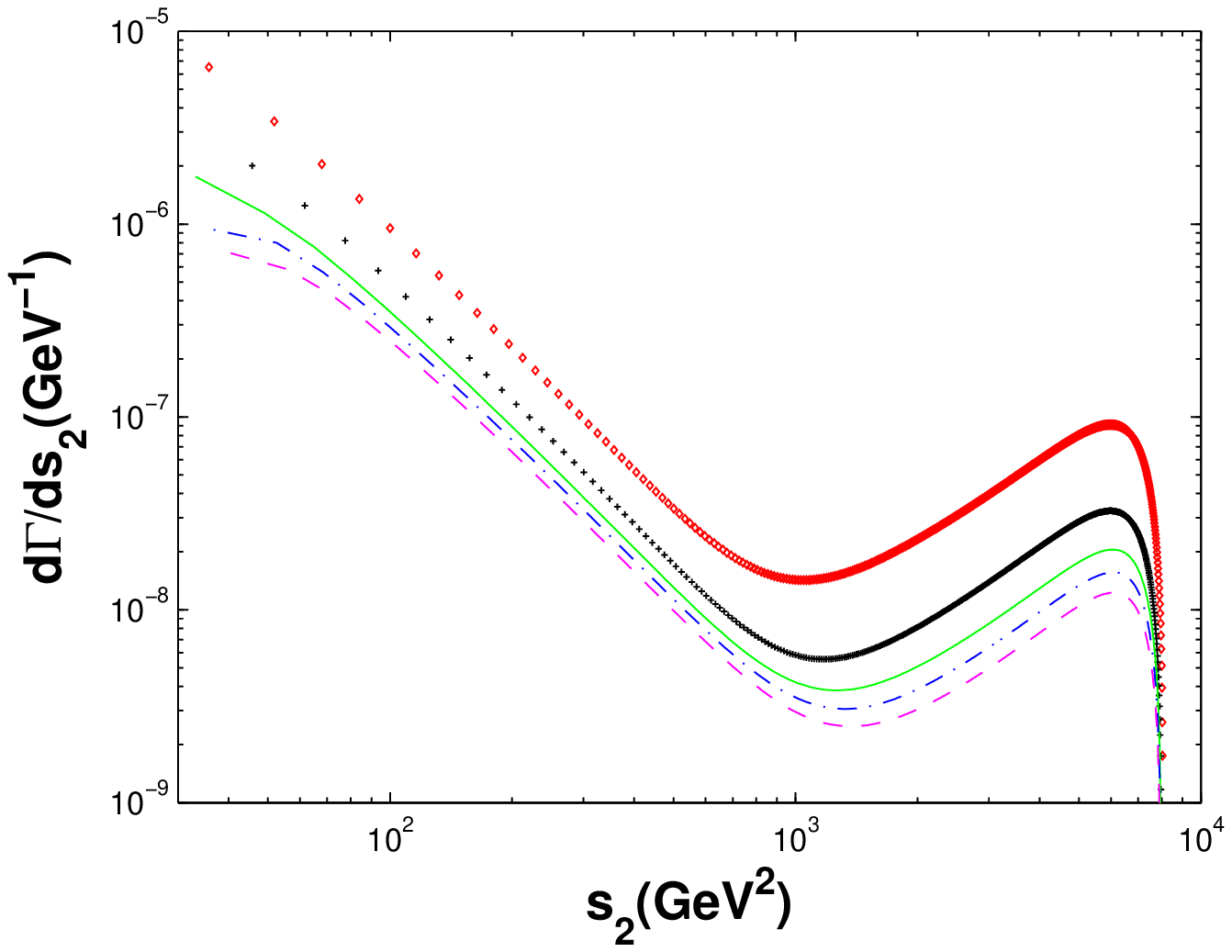}
\caption{(color online). Differential decay widths $d\Gamma/ds_1$ (Left) and $d\Gamma/ds_2$ (Right) for $Z^0\rightarrow |(c\bar{c})[n]\rangle +\bar{c}c$, where the diamond line, the dotted line, the solid line, the dash-dotted line, and the crossed line are for $|(c\bar{c})[n]\rangle$, where $n=1, \cdots, 5$, respectively. $|(c\bar{c})[n]\rangle$ stands for the summed decay width of $[n1^1S_0]$, $[n^3S_1]$, $[n^1P_1]$, and $[n^3P_J]$ ($J=0, 1, 2$) at the same $n$th level.} \label{zccds1ds2sum}
\end{figure}

\begin{figure}
\includegraphics[width=0.40\textwidth]{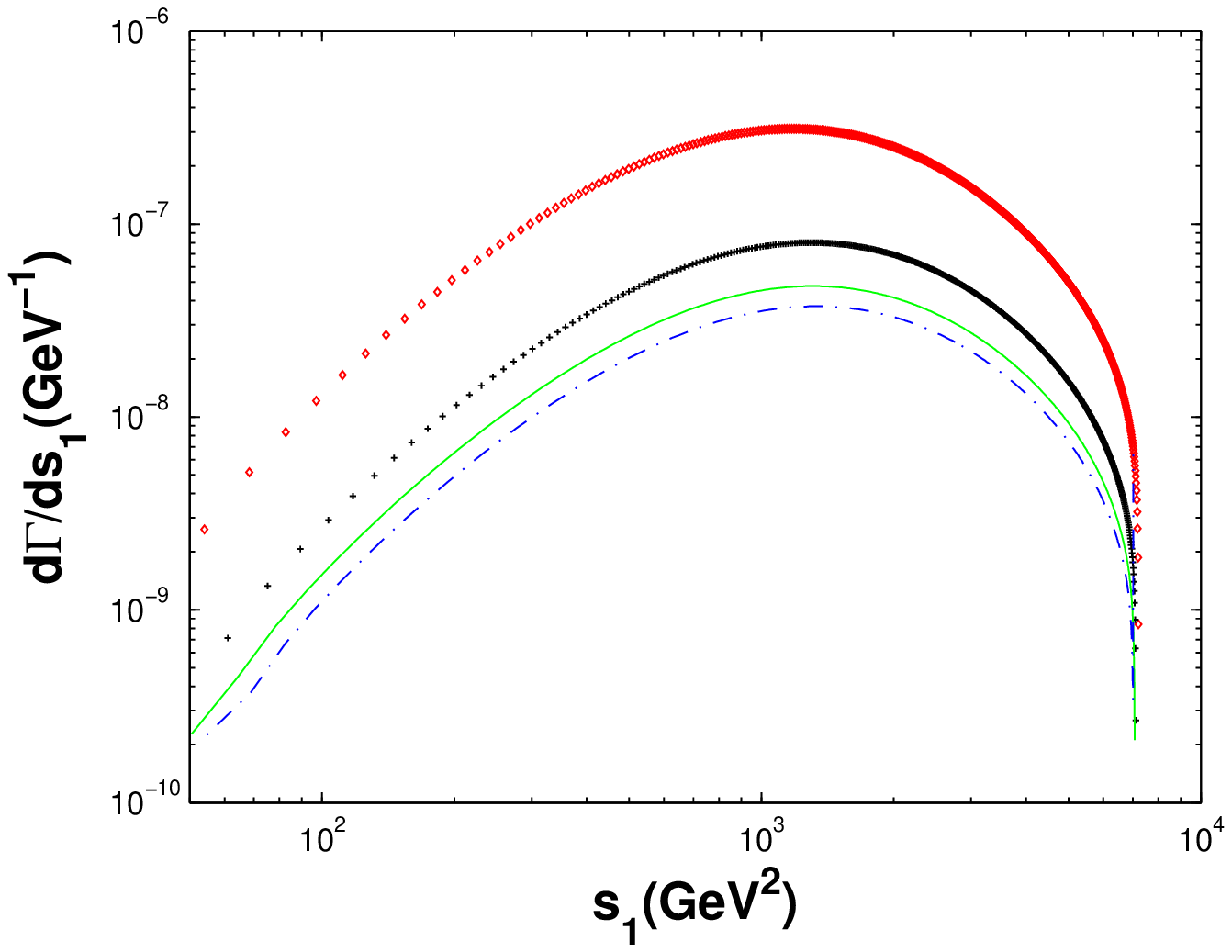}
\includegraphics[width=0.40\textwidth]{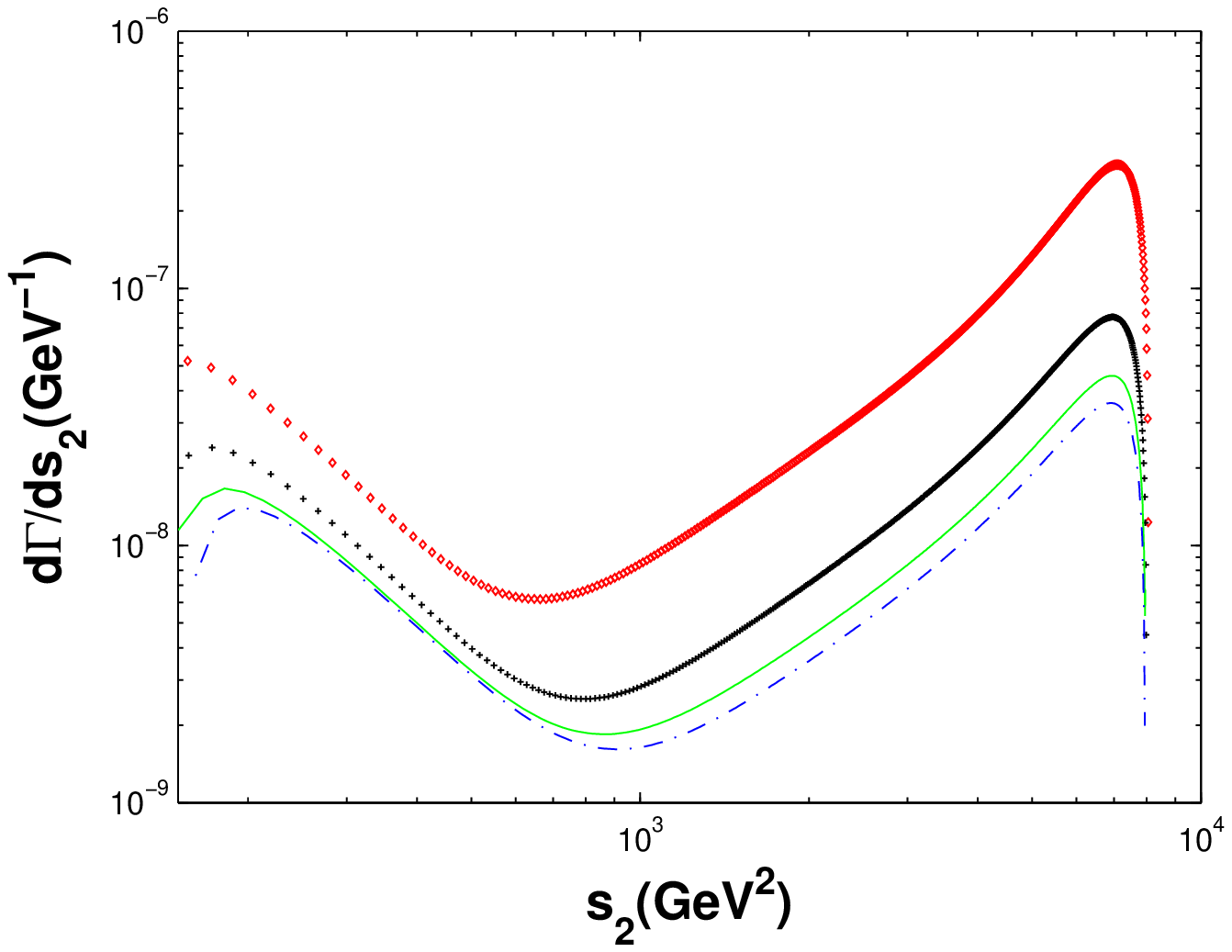}
\caption{(color online). Differential decay widths $d\Gamma/ds_1$ (Left) and $d\Gamma/ds_2$ (Right) for $Z^0\rightarrow |(b\bar{c})[n]\rangle +\bar{b}c$, where the diamond line, the dotted line, the solid line, and the dash-dotted line are for $|(b\bar{c})[n]\rangle$, where $n=1, \cdots, 4$, respectively. $|(b\bar{c})[n]\rangle$ stands for the summed decay width of $[n^1S_0]$, $[n^3S_1]$, $[n^1P_1]$, and $[n^3P_J]$ ($J=0, 1, 2$) at the same $n$th level.} \label{zbcds1ds2sum}
\end{figure}

\begin{figure}
\includegraphics[width=0.40\textwidth]{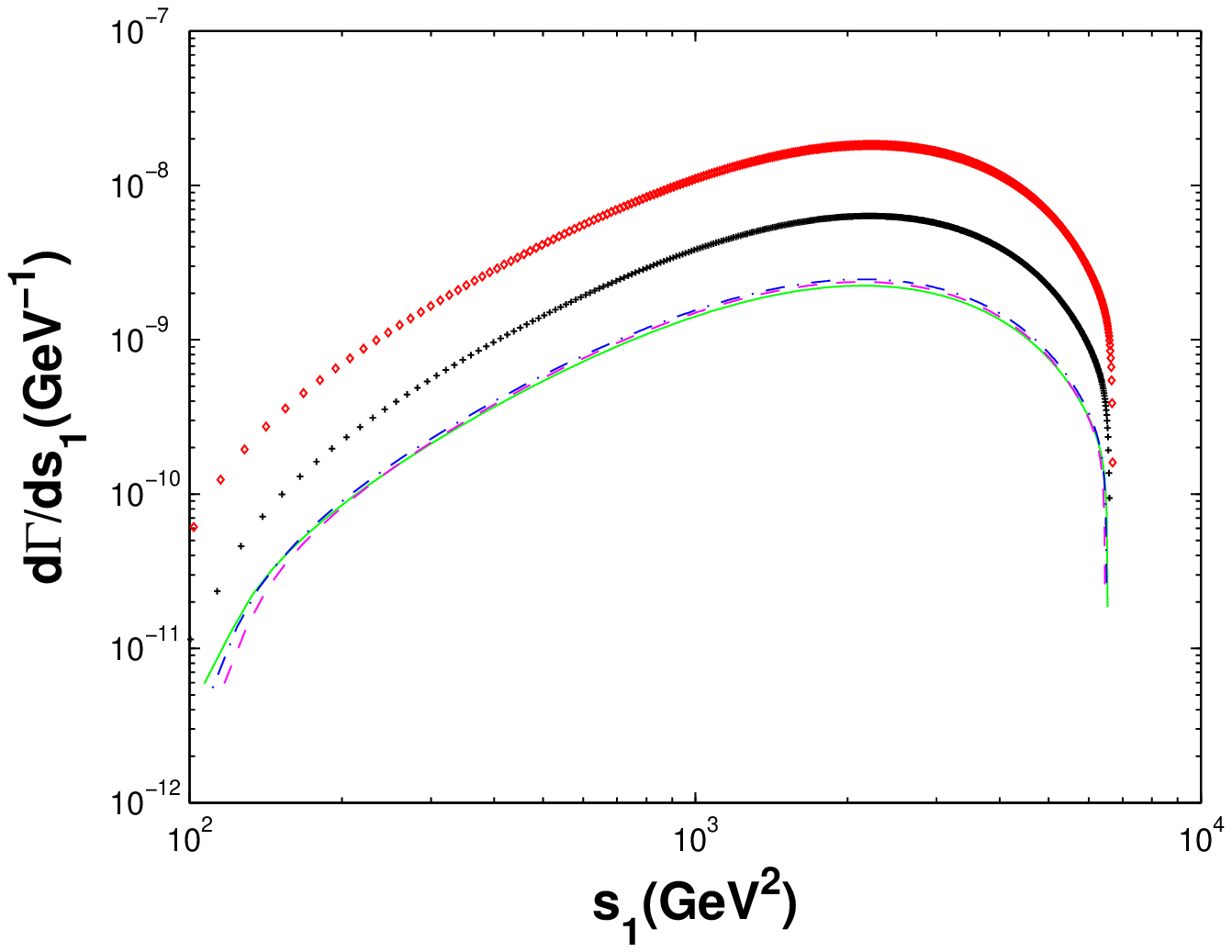}
\includegraphics[width=0.40\textwidth]{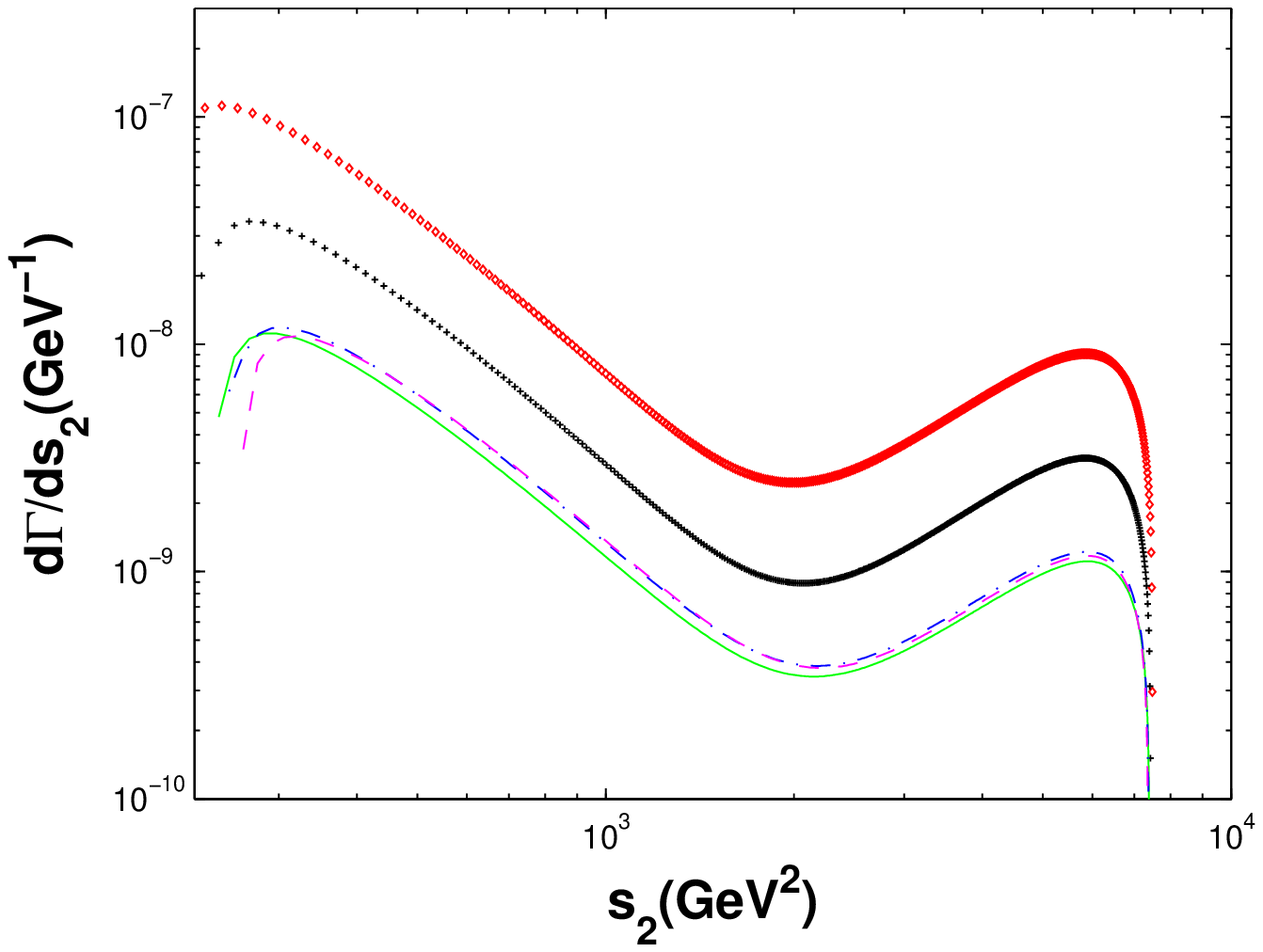}
\caption{(color online). Differential decay widths $d\Gamma/ds_1$ (Left) and $d\Gamma/ds_2$ (Right) for $Z^0\rightarrow |(b\bar{b})[n]\rangle +\bar{b}b$, where the diamond line, the dotted line, the solid line, the dash-dotted line, and the crossed line are for $|(b\bar{b})[n]\rangle$, where $n=1, \cdots, 5$, respectively. $|(b\bar{b})[n]\rangle$ stands for the summed decay width of $[n^1S_0]$, $[n^3S_1]$, $[n^1P_1]$, and $[n^3P_J]$ ($J=0, 1, 2$) at the same $n$th level.} \label{zbbds1ds2sum}
\end{figure}

\begin{figure}
\includegraphics[width=0.40\textwidth]{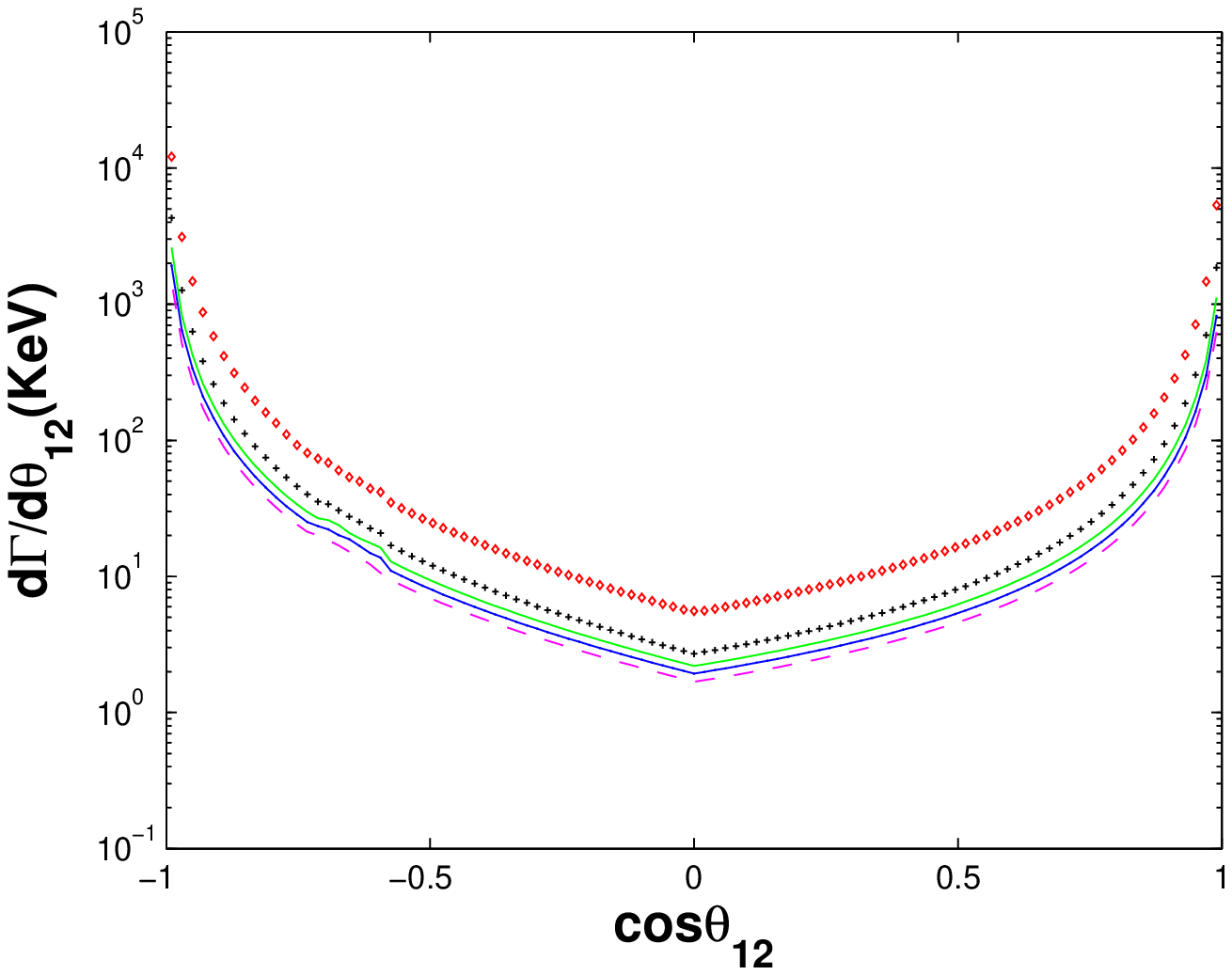}
\includegraphics[width=0.40\textwidth]{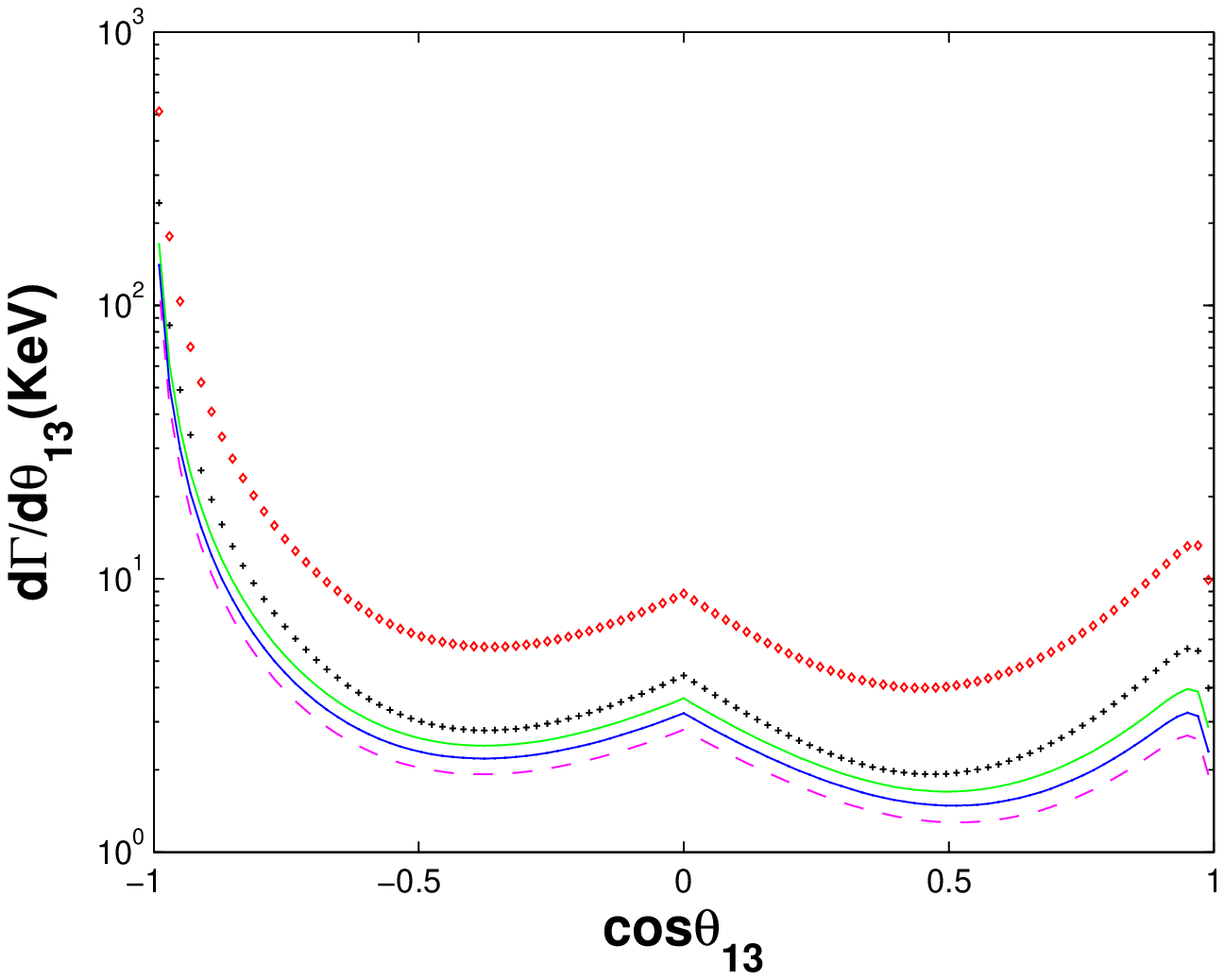}
\caption{(color online). Differential decay widths $d\Gamma/dcos\theta_{12}$ (Left) and $d\Gamma/dcos\theta_{13}$ (Right) for $Z^0\rightarrow |(c\bar{c})[n]\rangle +\bar{c}c$, where the diamond line, the dotted line, the solid line, the dash-dotted line, and the crossed line are for $|(c\bar{c})[n]\rangle$, where $n=1, \cdots, 5$, respectively. $|(c\bar{c})[n]\rangle$ stands for the summed decay width of $[n^1S_0]$, $[n^3S_1]$, $[n^1P_1]$, and $[n^3P_J]$ ($J=0, 1, 2$) at the same $n$th level.} \label{zccdcos12dcos13sum}
\end{figure}

\begin{figure}
\includegraphics[width=0.40\textwidth]{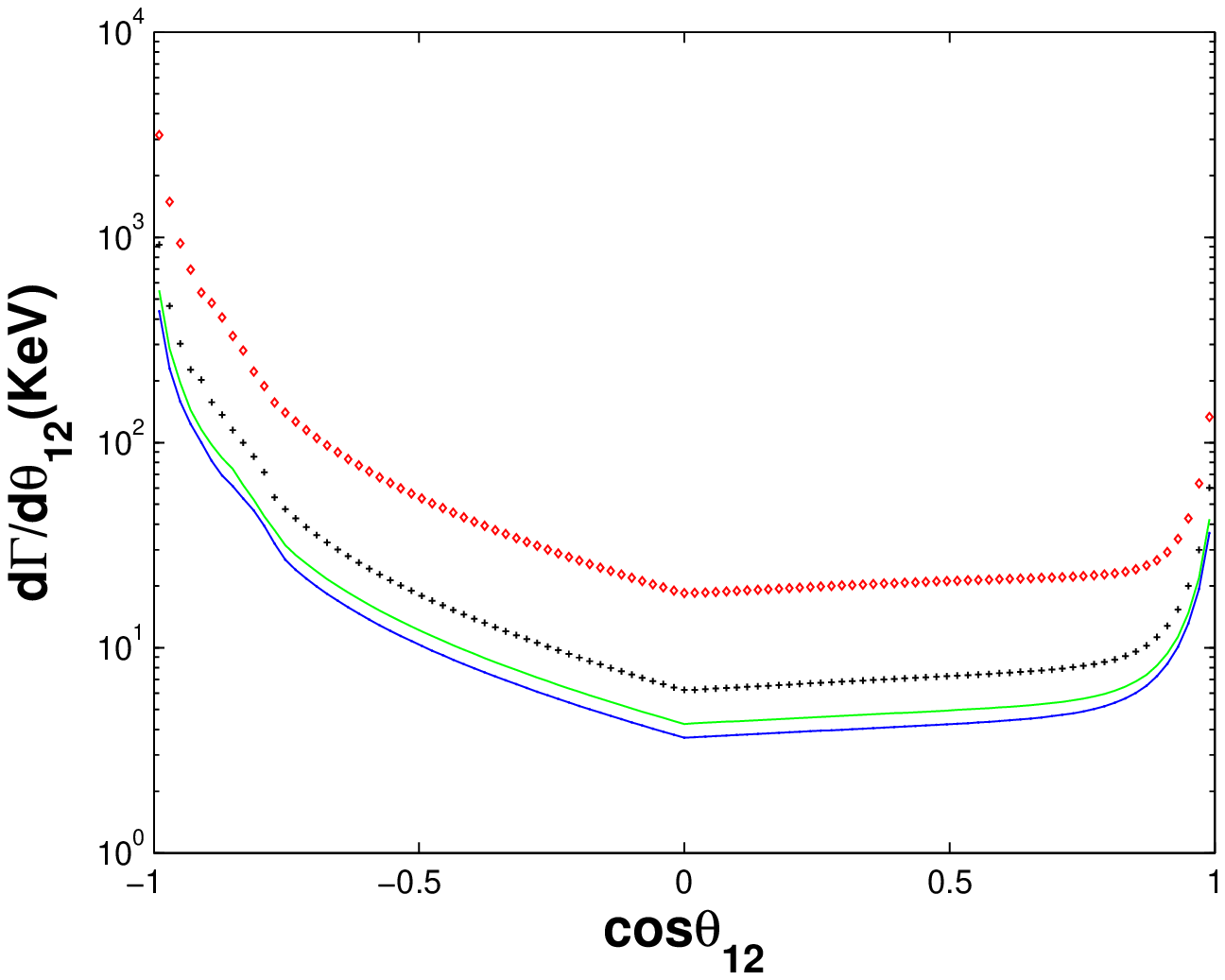}
\includegraphics[width=0.40\textwidth]{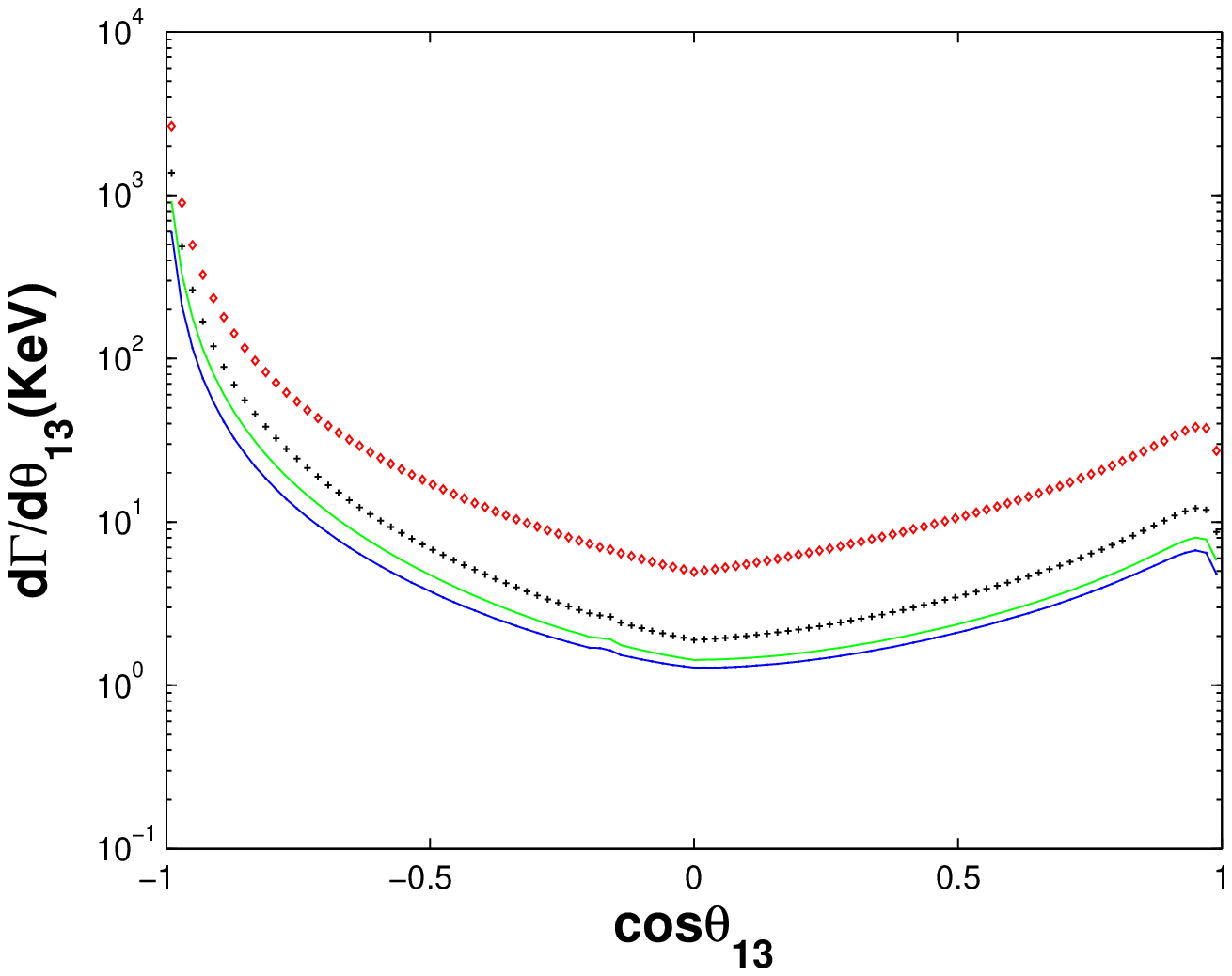}
\caption{(color online). Differential decay widths $d\Gamma/dcos\theta_{12}$ (Left) and $d\Gamma/dcos\theta_{13}$ (Right) for $Z^0\rightarrow |(b\bar{c})[n]\rangle +\bar{b}c$, where the diamond line, the dotted line, the solid line, and the dash-dotted line are for $|(b\bar{c})[n]\rangle$, where $n=1, \cdots, 4$, respectively. $|(b\bar{c})[n]\rangle$ stands for the summed decay width of $[n^1S_0]$, $[n^3S_1]$, $[n^1P_1]$, and $[n^3P_J]$ ($J=0, 1, 2$) at the same $n$th level.} \label{zbcdcos12dcos13sum}
\end{figure}
\end{center}
\end{widetext}

\begin{figure}
\includegraphics[width=0.40\textwidth]{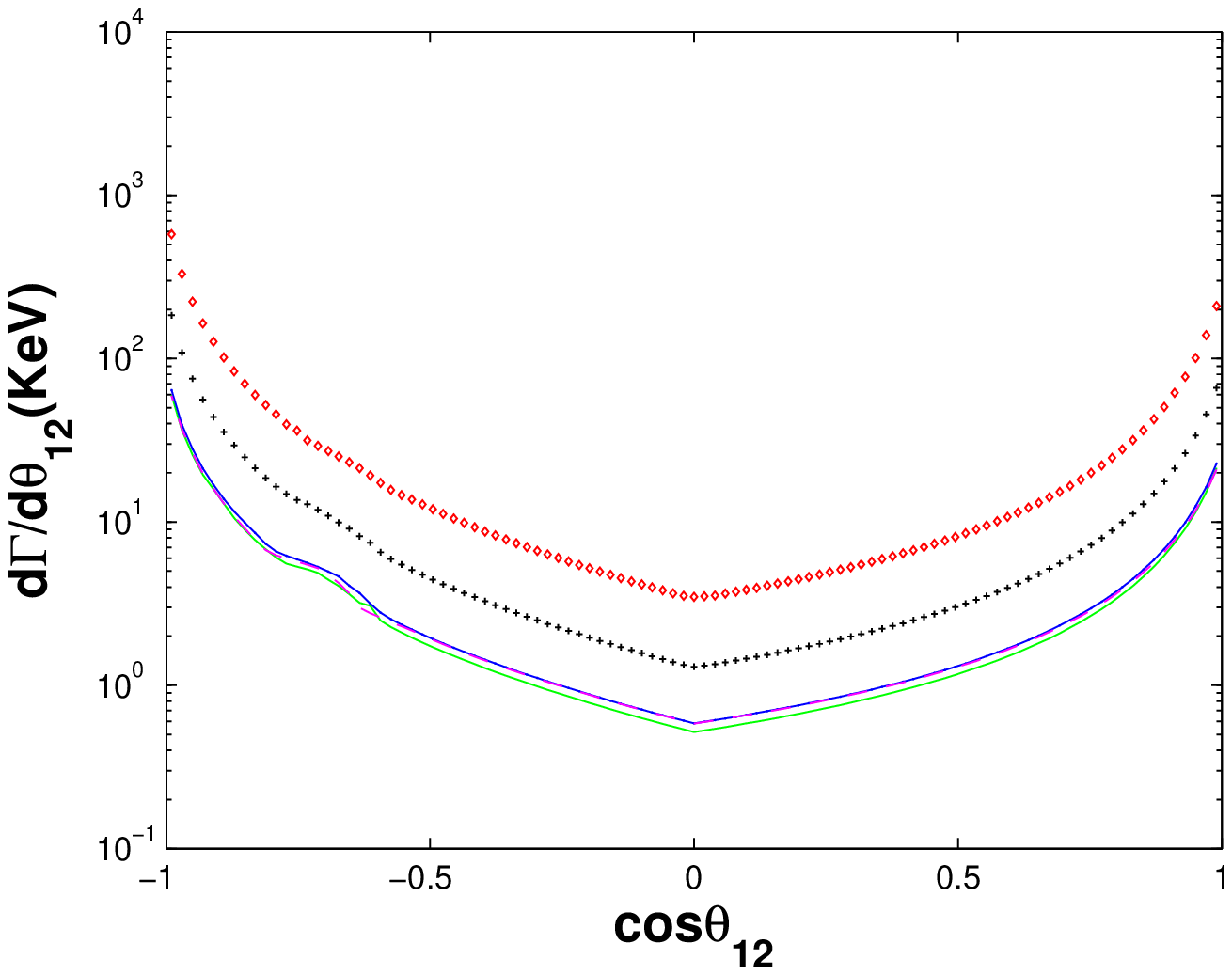}
\includegraphics[width=0.40\textwidth]{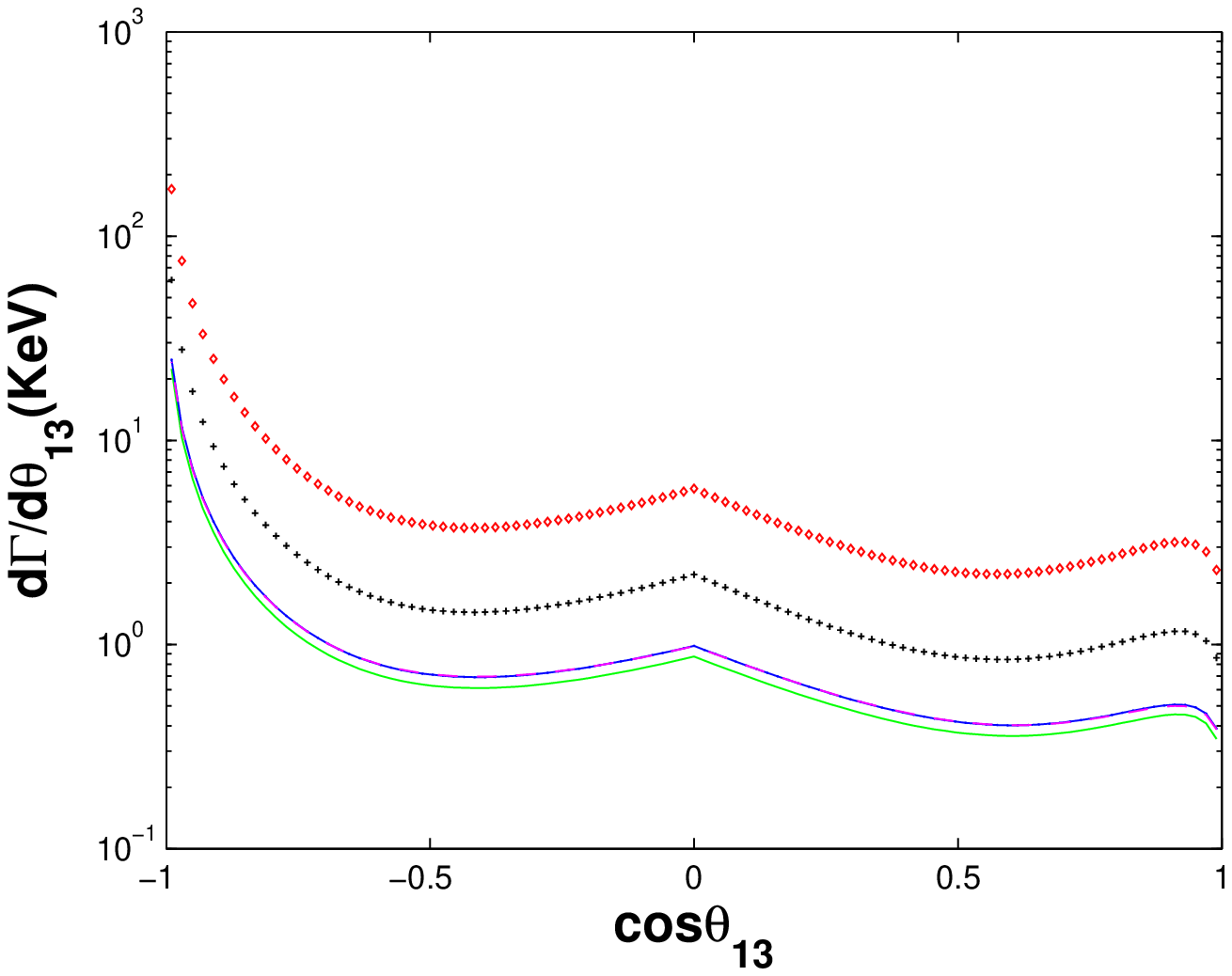}
\caption{(color online). Differential decay widths $d\Gamma/dcos\theta_{12}$ (Up) and $d\Gamma/dcos\theta_{13}$ (Down) for $Z^0\rightarrow |(b\bar{b})[n]\rangle +\bar{b}b$, where the diamond line, the dotted line, the solid line, the dash-dotted line, and the crossed line are for $|(b\bar{b})[n]\rangle$, where $n=1, \cdots, 5$, respectively. $|(b\bar{b})[n]\rangle$ stands for the summed decay width of $[n^1S_0]$, $[n^3S_1]$, $[n^1P_1]$, and $[n^3P_J]$ ($J=0, 1, 2$) at the same $n$th level.} \label{zbbdcos12dcos13sum}
\end{figure}

\begin{itemize}
\item For $|(c\bar{c})[n]\rangle$ quarkonium production through the decay channels $Z^0\rightarrow |(c\bar{c})[n]\rangle + \bar{c}c$, the total decay widths for all $2S$, $3S$, $4S$, $5S$, $6S$, $1P$, $2P$, $3P$, $4P$, and $5P$-wave states is $34.5\%$, $17.5\%$, $12.2\%$, $9.3\%$, $7.9\%$, $19.9\%$, $7.6\%$, $8.7\%$, $5.9\%$, and $6.2\%$ of the summed decay widths of $\eta_c$ and $J/\Psi$. Considering that the LHC runs at the center-of-mass energy $\sqrt{S}=14$ TeV and the ILC runs at the $Z^0$ pole energy with the luminosity ${\cal L}\propto 10^{34}cm^{-2}s^{-1}$, one expects that about $1.0\times10^9$ $Z^0$ events per year can be generated. Then we can estimate the charmonium events generated through $Z^0$ decays; i.e., $2.6\times10^5$  $|(c\bar{c})[1S]\rangle$, $8.9\times10^4$ $|(c\bar{c})[2S]\rangle$, $4.5\times10^4$ $|(c\bar{c})[3S]\rangle$, $3.2\times10^4$ $|(c\bar{c})[4S]\rangle$, $2.4\times10^4$ $|(c\bar{c})[5S]\rangle$, $2.0\times10^4$ $|(c\bar{c})[6S]\rangle$, $5.6\times10^4$ $|(c\bar{c})[1P]\rangle$, $2.0\times10^4$ $|(c\bar{c})[2P]\rangle$, $2.2\times10^4$ $|(c\bar{c})[3P]\rangle$, $1.5\times10^4$ $|(c\bar{c})[4P]\rangle$, and $1.6\times10^4$ $|(c\bar{c})[5P]\rangle$ quarkonium events per year can be obtained.
\end{itemize}

\begin{itemize}
\item For $|(b\bar{c})[n]\rangle$ quarkonium production via $Z^0$ boson semiexclusive decays, the total decay widths for all $2S$, $3S$, $4S$, $5S$, $1P$, $2P$, $3P$, and $4P$-wave states is $24.8\%$, $13.3\%$, $10.1\%$, $8.2\%$, $8.5\%$, $4.7\%$, $4.4\%$, and $3.9\%$ of the summed decay widths of $B_c$ and $B^*_c$ . At the LHC and ILC, $3.4\times10^5$  $|(b\bar{c})[1S]\rangle$, $8.4\times10^4$ $|(b\bar{c})[2S]\rangle$, $4.5\times10^4$ $|(b\bar{c})[3S]\rangle$, $3.4\times10^4$ $|(b\bar{c})[4S]\rangle$, $2.6\times10^4$ $|(b\bar{c})[5S]\rangle$, $2.9\times10^4$ $|(b\bar{c})[1P]\rangle$, $1.6\times10^4$ $|(b\bar{c})[2P]\rangle$, $1.5\times10^4$ $|(b\bar{c})[3P]\rangle$, and $1.3\times10^4$ $|(b\bar{c})[4P]\rangle$ quarkonium events per year can be obtained.
\end{itemize}

\begin{itemize}
\item For $|(b\bar{b})[n]\rangle$ quarkonium production via $Z^0$ boson semiexclusive decays, the total decay widths for all $2S$, $3S$, $4S$, $5S$, $6S$, $7S$, $1P$, $2P$, $3P$, $4P$, $5P$, and $6P$ wave states are about $34.4\%$, $9.7\%$, $10.8\%$ $10.1\%$, $9.2\%$, $8.0\%$, $10.9\%$, $4.3\%$, $3.6\%$, $3.8\%$, $3.8\%$, and $3.6\%$ of the summed decay widths of $\eta_b$ and $\Upsilon$. At the LHC and ILC, bout $2.4\times10^4$  $|(b\bar{b})[1S]\rangle$, $8.1\times10^3$ $|(b\bar{b})[2S]\rangle$, $2.3\times10^3$ $|(b\bar{b})[3S]\rangle$, $2.5\times10^3$ $|(b\bar{b})[4S]\rangle$, $2.4\times10^3$ $|(b\bar{b})[5S]\rangle$, $2.2\times10^3$ $|(b\bar{b})[6S]\rangle$, $1.9\times10^3$ $|(b\bar{b})[7S]\rangle$, and summed up, $7.1\times10^3$ $|(b\bar{b})[P]\rangle$ quarkonium events per year can be obtained.
\end{itemize}

To better illustrate the relative importance of different production channels, we present the differential distributions
$d\Gamma/ds_{1}$, $d\Gamma/ds_{2}$, $d\Gamma/dcos\theta_{12}$, and $d\Gamma/dcos\theta_{13}$ for the above-mentioned three processes in Figs. \ref{zccds1ds2sum},  \ref{zbcds1ds2sum},  \ref{zbbds1ds2sum},  \ref{zccdcos12dcos13sum}, \ref{zbcdcos12dcos13sum},  \ref{zbbdcos12dcos13sum}. We have used $|(Q\bar{Q'})[n]\rangle$ to represent the summed decay width of $(Q\bar{Q'})[n^1S_0]\rangle$,  $(Q\bar{Q'})[n^3S_1]\rangle$, $(Q\bar{Q'})[n^1P_1]\rangle$, and $(Q\bar{Q'})[n^3P_J]\rangle$ ($J=0, 1, 2$) at the same $n$th level in the figures, where $n=1, \cdots, 5$. These figures show explicitly that the excited Fock states, i.e., excited states of $|(Q\bar{Q'})[n]\rangle$ quarkioum $(n=2, \cdots, 5)$, can provide sizable contributions in comparison to the lower Fock state $|(Q\bar{Q'})[1^1S_0]\rangle$ or $|(Q\bar{Q'})[1^3S_1]\rangle$ in almost the entire kinematical region.

As all of the higher excited heavy quarkonium $nS$ and $nP$ Fock states almost decay to the ground spin-singlet $S$ wave state $|(Q\bar{Q'})[1^1S_0]\rangle$ via electromagnetic or hadronic interactions, we obtain the total decay width of $Z^0$ boson decay channels within the B.T. potential:
\begin{eqnarray}
\Gamma{(Z^0\to |(c\bar{c})[1^1S_0]\rangle +\bar{c}c)} &=&1476\;{\rm KeV} \label{cc} ,\\
\Gamma{(Z^0\to |(b\bar{c})[1^1S_0]\rangle +\bar{b}c)} &=&1485\;{\rm KeV} \label{bc},\\
\Gamma{(Z^0\to |(b\bar{b})[1^1S_0]\rangle +\bar{b}b)} &=&127.5\;{\rm KeV} \label{bb}.
\end{eqnarray}

Running at the center-of-mass energy $\sqrt{S}=14$ TeV at the LHC \cite{sc} and at the $Z^0$ pole energy at the ILC \cite{jsw,g} with luminosity $10^{34} cm^{-2} s^{-1}$, one may expect to produce about $1.0\times10^{9}$ $Z^0$ boson per year. Then we can estimate the event number of $|(Q\bar{Q'})\rangle$ quarkonium production through $Z^0$ boson decays, i.e., about $5.9\times10^5$ $|(c\bar{c})[n]\rangle$ quarkonium events, $6.0\times10^5$ $|(b\bar{c})[n]\rangle$ (or $|(c\bar{b})[n]\rangle$) quarkonium events, $5.1\times10^4$ $|(b\bar{b})[n]\rangle$ quarkonium events per year. Bearing in mind the situation pointed out here and the possible upgrade for the LHC (SLHC, DLHC, etc.~\cite{ab}), and the newly purposed $Z$ factory with luminosity $10^{36} cm^{-2} s^{-1}$ \cite{mz}, the possibility to study $|(Q\bar{Q'})[n]\rangle$ quarkonium via $Z^0$ boson decays is worth serious consideration.

\subsection{Decay widths under five potential models}

In this subsection, we discuss the uncertainties for the $|(Q\bar{Q'})[n]\rangle$ quarkonium production through
$Z^0$ boson decays. For the present calculation, their main uncertainty sources include the nonperturbative bound state
matrix elements, the renormalization scale $\mu_R$, and the constituent quark masses $m_b$ and $m_c$. These parameters are the main uncertainty source for estimating heavy $|(Q\bar{Q'})[n]\rangle$-quarkonium production. In the present, we discuss the decay widths of $|(Q\bar{Q'})[n]\rangle$ quarkonium production through $Z^0$ boson decays under the five potential models , i.e., the B.T. potential \cite{lx,pot2}, the John L. Richardson potential (J. potential)~\cite{lx,jlr}, the K. Igi and S. Ono potential (I.O. potential)~\cite{lx,kso,sr}, the Yu-Qi Chen and Yu-Ping Kuang potential(C.K. potential)~\cite{lx,pot5,sr}, and Coulomb-plus-linear potential (Cor. potential)~\cite{lx,pot1} in detail.

\begin{widetext}
\begin{center}
\begin{table}
\caption{Decay widths (in unit keV) for the $|(c\bar{c})[n]\rangle$ quarkonium production channel $Z^0\rightarrow |(c\bar{c})[n]\rangle+\bar{c}c$, whose bound-state parameters are adopted in the five potential models in Ref. \cite{lx}.}
\begin{tabular}{|c||c|c|c|c|c|c|c|}
\hline
~~~~&B.T. nf=3~/~4 \cite{pot2} &J. nf=3 / 4 \cite{jlr}&I.O. nf=3 / 4 \cite{kso}&C.K. nf=3 / 4 \cite{pot5}&Cor. \cite{pot1}\\
\hline\hline
$[n]=[1^1S_0]$ &313.0~/~298.5&142.5~/~127.0&71.95~/~76.28&92.45~/~101.2&~124.2~\\
\hline
$[n]=[1^3S_1]$ &325.0~/~309.9&148.0~/~131.8&74.71~/~79.20&95.99~/~105.1&~128.9~\\
\hline
$[n]=[2^1S_0]$ &108.0~/~90.83&70.60~/~60.78&36.67~/~38.07&41.01~/~43.55&~59.38~\\
\hline
$[n]=[2^3S_1]$ &112.3~/~94.45&73.41~/~63.20&38.13~/~39.59&42.64~/~45.28&~61.74~\\
\hline
$[n]=[3^1S_0]$ &54.67~/~49.76&55.57~/~47.17&29.23~/~30.13&31.48~/~32.95&~46.32~\\
\hline
$[n]=[3^3S_1]$ &56.93~/~51.82~&57.87~/~49.12&30.43~/~31.37&32.78~/~34.31&~48.23~\\
\hline
$[n]=[4^1S_0]$ &38.25~/~38.11&46.61~/~39.21&24.65~/~25.32&26.04~/~27.10&~38.78~\\
\hline
$[n]=[4^3S_1]$ &39.86~/~39.71&48.57~/~40.86&25.69~/~26.39&27.14~/~28.24&~40.41~\\
\hline
$[n]=[5^1S_0]$ &28.90~/~27.12&37.76~/~31.61&20.08~/~20.58&20.93~/~21.67&~31.41~\\
\hline
$[n]=[5^3S_1]$ &30.13~/~28.27&39.37~/~32.95&20.93~/~21.46&21.82~/~22.59&~32.75~\\
\hline
$[n]=[6^1S_0]$ &24.59~/~24.62&34.38~/~28.68&18.35~/~18.79&19.00~/~19.58&~28.72~\\
\hline
$[n]=[6^3S_1]$ &25.66~/~25.70&35.87~/~29.93&19.15~/~19.61&19.82~/~20.43&~29.97~\\
\hline
$[n]=[1P]$ &126.8~/~129.6&67.75~/~53.17&20.88~/~22.45&29.15~/~31.90&~35.84~\\
\hline
$[n]=[2P]$&48.55~/~49.85&66.97~/~51.36&21.46~/~22.54&27.74~/~30.13&~36.63~\\
\hline
$[n]=[3P]$ &55.29~/~54.01&62.44~/~47.43&20.29~/~21.29&25.29~/~27.29&~34.86~\\
\hline
$[n]=[4P]$ &48.26~/~48.99&58.49~/~44.13&19.22~/~20.15&23.36~/~25.11&~33.07~\\
\hline
$[n]=[5P]$ &39.74~/~40.94&55.27~/~41.50&18.40~/~19.11&21.90~/~23.42&~29.95~\\
\hline
Sum. &1476~~/~1402&1138~~/~919.9&510.2~/~532.3&598.5~/~639.9&~841.2~\\
\hline
\end{tabular}
\label{tabrpe}
\end{table}

\begin{table}
\caption{Decay widths (in unit keV) for $|(b\bar{c})[n]\rangle$ quarkonium production channel $Z^0\rightarrow |(b\bar{c})[n]\rangle+\bar{b}c$, whose bound-state parameters are adopted in the five potential models in Ref. \cite{lx}.}
\begin{tabular}{|c||c|c|c|c|c|c|c|}
\hline
~~~~&B.T. nf=3~/~4 \cite{pot2} &J. nf=3 / 4 \cite{jlr}&I.O. nf=3 / 4 \cite{kso}&C.K. nf=3 / 4 \cite{pot5}&Cor. \cite{pot1}\\
\hline\hline
$[n]=[1^1S_0]$ &342.2~/~356.5&179.7~/~162.7&552.3~/~467.9&116.0~/~128.7&~158.6~\\
\hline
$[n]=[1^3S_1]$ &492.4~/~513.0&258.6~/~234.0&794.8~/~673.3&166.9~/~185.2&~228.2~\\
\hline
$[n]=[2^1S_0]$ &87.71~/~61.67&79.68~/~69.17&95.74~/~86.43&46.17~/~49.22&~70.36~\\
\hline
$[n]=[2^3S_1]$ &119.5~/~84.02&108.6~/~94.24&130.4~/~117.8&62.91~/~67.06&~95.86~\\
\hline
$[n]=[3^1S_0]$ &47.22~/~42.38&58.05~/~49.57&45.61~/~41.58&32.71~/~34.32&~51.32~\\
\hline
$[n]=[3^3S_1]$ &63.46~/~56.96&78.02~/~66.62&61.29~/~55.88&43.96~/~46.12&~68.97~\\
\hline
$[n]=[4^1S_0]$ &36.11~/~36.56&45.82~/~38.74&26.57~/~24.42&25.49~/~26.51&~40.71~\\
\hline
$[n]=[4^3S_1]$ &48.02~/~48.62&60.94~/~51.51&35.33~/~32.48&33.90~/~35.25&~54.14~\\
\hline
$[n]=[5^1S_0]$ &29.22~/~31.83&41.06~/~34.47&19.18~/~17.69&22.69~/~23.44&~36.41~\\
\hline
$[n]=[5^3S_1]$ &38.96~/~42.44&54.75~/~45.96&25.58~/~23.59&30.25~/~31.26&~48.54~\\
\hline
$[n]=[1P]$ &71.16~/~103.9&56.74~/~45.47&78.72~/~64.70&25.55~/~28.71&~30.09~\\
\hline
$[n]=[2P]$&39.45~/~34.40&54.12~/~42.37&38.11~/~32.35&24.62~/~27.30&~29.98~\\
\hline
$[n]=[3P]$ &36.94~/~39.27&47.78~/~36.94&21.08~/~18.19&19.76~/~21.58&~27.21~\\
\hline
$[n]=[4P]$ &32.70~/~36.91&45.85~/~35.12&14.46~/~12.59&19.82~/~21.57&~26.54~\\
\hline
Sum. &1485~~/~1488&1169~~/~1007&1939~~/~1669&671.5~/~726.2&~966.9~\\
\hline
\end{tabular}
\label{tabrpf}
\end{table}

\begin{table}
\caption{Decay widths (in unit keV) for $|(b\bar{b})[n]\rangle$-quarkonium production channel $Z^0\rightarrow |(b\bar{b})[n]\rangle+\bar{b}b$, whose bound-state parameters are adopted in the five potential models in Ref. \cite{lx}.}
\begin{tabular}{|c||c|c|c|c|c|c|c|}
\hline
~~~~&B.T. nf=4~/~5 \cite{pot2} &J. nf=4 / 5 \cite{jlr}&I.O. nf=4 / 5 \cite{kso}&C.K. nf=4 / 5 \cite{pot5}&Cor. \cite{pot1}\\
\hline\hline
$[n]=[1^1S_0]$ &28.48~/~24.61&12.51~/~9.827&17.55~/~15.29&9.314~/~10.69&~16.07~\\
\hline
$[n]=[1^3S_1]$ &31.83~/~27.51&14.04~/~10.98&19.61~/~17.09&10.41~/~11.95&~17.96~\\
\hline
$[n]=[2^1S_0]$ &9.537~/~10.49&5.861~/~4.083&4.894~/~4.262&3.934~/~4.230&~6.745~\\
\hline
$[n]=[2^3S_1]$ &10.74~/~11.81&6.601~/~4.598&5.512~/~4.800&4.431~/~4.763&~7.596~\\
\hline
$[n]=[3^1S_0]$ &2.737~/~6.093&4.286~/~2.845&2.585~/~2.252&2.797~/~2.930&~4.916~\\
\hline
$[n]=[3^3S_1]$ &3.095~/~6.890&4.846~/~3.218&2.923~/~2.546&3.163~/~3.313&~5.559~\\
\hline
$[n]=[4^1S_0]$ &3.039~/~3.476&3.578~/~2.314&1.707~/~1.488&2.316~/~2.392&~4.109~\\
\hline
$[n]=[4^3S_1]$ &3.446~/~3.941&4.057~/~2.624&1.936~/~1.687&2.626~/~2.712&~4.659~\\
\hline
$[n]=[5^1S_0]$ &2.839~/~2.377&3.074~/~1.958&1.218~/~1.063&1.983~/~2.029&~3.541~\\
\hline
$[n]=[5^3S_1]$ &3.231~/~2.705&3.499~/~2.228&1.386~/~1.210&2.256~/~2.310&~4.030~\\
\hline
$[n]=[6^1S_0]$ &2.581~/~2.251&2.767~/~1.742&0.943~/~0.824&1.783~/~1.814&~3.189~\\
\hline
$[n]=[6^3S_1]$ &2.944~/~2.568&3.157~/~1.987&1.075~/~0.939&2.034~/~2.069&~3.638~\\
\hline
$[n]=[7^1S_0]$ &2.256~/~2.192&2.540~/~1.585&0.761~/~0.665&1.636~/~1.656&~2.927~\\
\hline
$[n]=[7^3S_1]$ &2.579~/~2.506&2.903~/~1.812&0.870~/~0.761&1.871~/~1.893&~3.346~\\
\hline
$[n]=[1P]$ &6.560~/~5.554&1.836~/~0.986&1.301~/~1.021&1.241~/~1.501&~1.360~\\
\hline
$[n]=[2P]$&2.608~/~4.812&1.980~/~0.987&0.890~/~0.700&1.221~/~1.427&~1.538~\\
\hline
$[n]=[3P]$ &2.190~/~3.411&2.084~/~0.996&0.675~/~0.531&1.232~/~1.412&~1.666~\\
\hline
$[n]=[4P]$ &2.300~/~2.163&2.031~/~0.946&0.500~/~0.396&1.170~/~1.326&~1.662~\\
\hline
$[n]=[5P]$ &2.312~/~2.053&2.033~/~0.929&0.402~/~0.319&1.151~/~1.292&~1.691~\\
\hline
$[n]=[6P]$ &2.175~/~2.099&2.034~/~0.915&0.335~/~0.266&1.135~/~1.266&~1.711~\\
\hline
Sum. &127.5~/~148.4&85.72~/~57.56&67.07~/~58.11&57.70~/~62.98&~~97.91~~\\
\hline
\end{tabular}
\label{tabrpg}
\end{table}
\end{center}
\end{widetext}

The decay widths for $|(Q\bar{Q'})[n]\rangle$ quarkonium production under the five potential models with the different number of flavor quarks $n_f$ presented in Tables \ref{tabrpe}, \ref{tabrpf}, and \ref{tabrpg}, where $[nP]$ represent the summed decay widths of $(Q\bar{Q'})[n^1P_1]\rangle$ and $(Q\bar{Q'})[n^3P_J]\rangle$ [with $J=(0, 1, 2)$] at the same $n$th level. The decay widths for the other four models are consistent with each other: taking the B.T. model decay width as the center value, for the channel $Z^0\rightarrow |(c\bar{c})[n]\rangle+\bar{c}c$, we obtain the uncertainty $(^{+0\%}_{-65\%})$ for $n_f=3$ and $(^{+0\%}_{-62\%})$ for $n_f=4$, where the lower value is for the I.O. potential; for the channel $Z^0\rightarrow |(b\bar{c})[n]\rangle+\bar{b}c$, we obtain the uncertainty $(^{+31\%}_{-55\%})$ for $n_f=3$ and $(^{+12\%}_{-51\%})$ for $n_f=4$, where the upper value is for the I.O. potential and the lower value is for the C.K. potential; for the channel $Z^0\rightarrow |(b\bar{b})[n]\rangle+\bar{b}b$, we obtain the uncertainty $(^{+0\%}_{-55\%})$ for $n_f=3$ and $(^{+0\%}_{-58\%})$ for $n_f=4$, where the lower value is for the C.K. potential.

\section{Conclusions}

In this paper, we have made a detailed study on the $|(c\bar{c})[n]\rangle$, $|(b\bar{c})[n]\rangle$, $|(b\bar{b})[n]\rangle$-quarkonium via $Z^0$ boson semiexclusive decays under the NRQCD framework, i.e., $Z^0\to |(c\bar{c})[n]\rangle +\bar{c}c$, $Z^0\to |(b\bar{c})[n]\rangle +\bar{b}c$, and $Z^0\to |(b\bar{b})[n]\rangle +\bar{b}b$, where $[n]$ stands for $[n^1S_0]$, $[n^3S_1]$, $[n^1P_1]$, and $[n^3P_J]$, ($n=1, \cdots, 7; J=[0 ,1 , 2]$). To provide the analytical expressions as simply as possible, we have adopted the improved trace technology to derive Lorentz-invariant expressions for $Z^0$ boson decay processes at the amplitude level. Such a calculation technology will be very helpful for dealing with processes with massive spinors.

Numerical results show that high excited states of $|(Q\bar{Q})[n]\rangle$ quarkonium in addition to the ground $1S$ wave states can also provide sizable contributions to heavy quarkonium production through $Z^0$ boson decays, so one needs to take the excited wave states into consideration for a sound estimation. If all the excited states decay to the ground state $|(Q\bar{Q})[1^1S_0]\rangle$ with $100\%$ efficiency via electromagnetic or hadronic interactions, we can obtain the total decay width for $|(Q\bar{Q'})\rangle$ quarkonium production through $Z^0$ boson decays as shown by Eqs.~(\ref{cc})-(\ref{bb}). At the LHC at runs with center-of-mass energy $\sqrt{S}=14$ TeV and at the ILC at the $Z^0$ pole energy with the luminosity ${\cal L}\propto 10^{34}cm^{-2}s^{-1}$, due to the high collision energy and high luminosity, sizable heavy quarkonium events can be produced through $Z^0$ boson decays; i.e., about $5.9~\times10^{5}$ $(c\bar{c})$, $6.0~\times10^{5}$ $(b\bar{c})$ [or$(c\bar{b})$], $5.1~\times10^{4}$ , $(b\bar{b})$ events per year can be obtained. At the newly purposed $Z$ factory with the high luminosity ${\cal L}\propto 10^{36}cm^{-2}s^{-1}$, the $|(Q\bar{Q'})\rangle$ quarkonium through $Z^0$ boson decays will be more abundantly produced. Therefore, one needs to take these higher excited states into consideration for a sound evaluation.

\hspace{2cm}

{\bf Acknowledgements}: We thank Professor Xing-Gang Wu for useful discussions.

\end{document}